# Ultrasoft transient X-ray emission from AGN

K. Mannheim[1], D. Grupe[1], K. Beuermann[1,2], H.-C. Thomas[3], and H.H. Fink[2]

[1] Universitäts-Sternwarte, Geismarlandstraße 11, D-37083 Göttingen, Germany
[2] Max–Planck–Institut für Extraterrestrische Physik, Giessenbachstrasse, D-85740 Garching, Germany
[3] Max–Planck–Institut für Astrophysik, Giessenbachstrasse, D-85740 Garching, Germany

**Abstract.** We report some remarkable and unexpected results of optical/UV/X-ray follow-up observations of bright soft X-ray selected AGN from the ROSAT All-Sky-Survey. The majority of these AGN are rather anonymous Seyfert galaxies, mostly of the narrow-line Seyfert 1 subtype (nlSy1). We confirm the well-known X-ray variability by factors of $\sim$ few. However, we also found strikingly different variability patterns: (i) a drop in the PSPC count rate by a factor of $\sim$ 400 in WPVS007, (ii) a bolometrically dominant soft X-ray component decreasing in flux by a factor $\sim$ 100 in IC 3599 accompanied by a decrease in optical emission line fluxes, and (iii) a drastic X-ray spectral change in RX J0134-42 from ultrasoft to a typical hard Seyfert spectrum. These dramatic variations occur within a (few) year(s), implying that the accretion flow in the immediate vicinity of the central black hole must have undergone a major change. We discuss possible physical explanations such as accretion disk instabilities or the tidal disruption of stars.

As part of our optical identification program of soft bright AGN discovered in the ROSAT All-Sky Survey (RASS), we investigated 29 sources (mostly nlSy1s) which received individual pointed PSPC exposure 1-2 years after the RASS. Selection criteria for the sources were (i) RASS count rate CR > 0.5 counts s$^{-1}$ and (ii) PSPC hardness ratio HR < 0 (corresponding to $\alpha_x \gtrsim 1.6$). Data reprocessing changed some count rates below the original cut (Grupe et al. 1996).

As shown in Fig.1, variability by factors of $\sim$few seems to be a common property. Large-amplitude variability is also present in well-sampled individual spectra of nlSy1 galaxies on time scales < 1 day (Boller et al. 1996, Marshall et al. 1996). Two sources show even more radical behavior: WPVS007 dropped in count rate by a factor $\sim$ 400 (Grupe et al. 1995), IC3599 by a factor of $\sim$ 100 (Brandt et al. 1995, Grupe et al. 1995). The sources had ultrasoft X-ray spectra in the RASS rivaled in steepness only by AM Her stars or hot white dwarfs. The spectral shape did not vary much during the change in count rate (Fig.2), although this conclusion is hampered in the case of WPVS007 by the lack of a well-determined spectrum from the pointed observation. Although showing a rather constant count rate, RX J0134-42 has changed its spectral shape from ultrasoft to a source with a hard X-ray spectrum rather typical for AGN (Fig.2). The fact that the count rate remained almost constant implies that the ultrasoft component must have decreased by a very large factor ($\gtrsim$ 10, see Fig.3). WPVS007 with its ultrasoft spectrum would not have shown up in a second all-sky survey and may therefore be coined an *'ultrasoft transient'*. On the other hand, RX J0134-42 with hardly any hard X-rays during the RASS would appear as a *'hard transient'* in a hard X-ray survey. Whether or not this 'transience' has any ramification to galactic transient X-ray sources is unknown at present. It could well be true, that the ultrasoft X-ray emission in the bright and soft sample is generally transient, and that the RASS represents a bias towards 'active' sources. The activity could either be on a very short time scale, as suggested by recent ASCA observations of IRAS 13224-3809 (Otani et al., this volume) which showed burst-like variations by a factor of $\sim 30$ over a period of 2 days, or that the components remain active over a few years. The latter behavior has been observed in the Seyfert galaxy E1615+061 by HEAO 1, Einstein and EXOSAT (Piro et al. 1988). Thermal instabilities in an accretion disk could be responsible for the observed ultrasoft transient emission. However, the viscous time scale (relevant for the decay of a hot state of the accretion disk) can be as short as a year only for extreme parameters: viscosity parameter $\alpha \sim 1$, black hole mass $M \sim 10^6 M_\odot$ (would grow rapidly due to accretion) and a localization of the instability in the very inner part of the disk $r \lesssim 10 r_G$ ($r_G = 2GM/c^2$). More luminous objects exceeding the Eddington luminosity of a $10^6 M_\odot$ black hole, such as RX J0134-42, are certainly *not* expected to change their central temperature within a year. Rapid spectral variability could also be due to changes in coronal plasma properties on a dynamical time scale, e.g. due to mass ejection into a disk wind (Mannheim et al. 1995). If the ultrasoft X-ray components are only short-lived ($\sim$days), their effect on the broad emission line gas surrounding the AGN could be negligible – unless the recurrence time is also very short. If the ultrasoft components persist for a time of the order of the light travel time across the broad emission line region ($\lesssim$ year), they would certainly change the ionization equilibrium in the emission line clouds. Indeed, we have discovered a 'line echo' in the highly ionized iron line flux from IC3599 soon after its RASS detection (Grupe et al. 1995). Comparing the Balmer line fluxes in our spectra with those of the spectrum shown in Brandt et al. (1995), we conclude that they, too, have decreased. However, IC3599 is different from the other transients: its ultrasoft component peaks in the PSPC band (blackbody-like with $kT \sim 100$ eV) and seems to dominate the bolometric luminosity (absence of strong FIR emission). The X-ray lightcurve shows an asymptotic decrease reminiscent of a powerful, possibly super-Eddington, *outburst* caused by a dramatic accretion event, such as the tidal disruption of a star by a black hole of mass $\sim 10^6 M_\odot$ (Rees 1988,1990). From Fig.4 we can estimate that the event could have occurred in late June 1990. Unfortunately, the short sampling time of the RASS detection does not allow for a discovery of the expected periodicity in the X-ray emission of the initially elliptical accretion disk formed by the debris of the disrupted star.

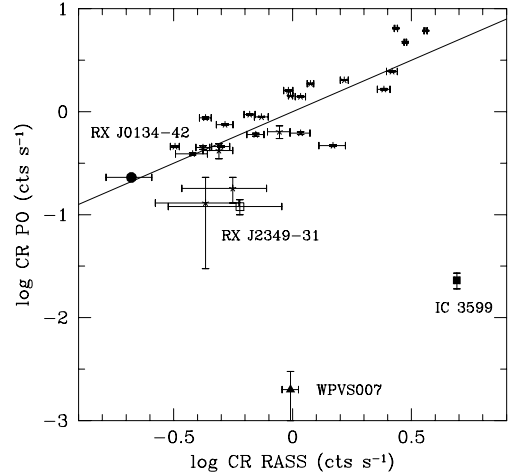

**Fig. 1.** RASS vs. pointed observation PSPC count rate for bright and soft AGN. Solid line indicates constant count rate

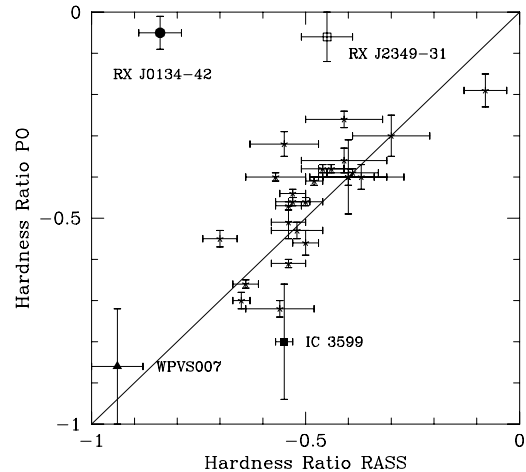

**Fig. 2.** Correspondingly, for the PSPC hardness ratio

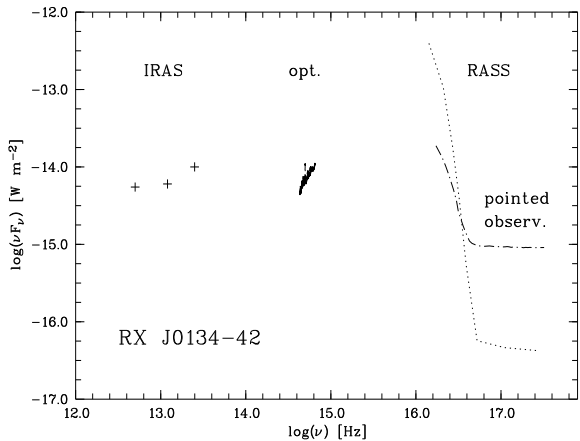

**Fig. 3.** Spectral energy distribution of RX J0134-42. Note the extremely blue optical spectrum (quasi-contemporaneous with the pointed PSPC observation) and the drastic change in the PSPC spectrum between RASS and pointed observation two years later. The uncertainty in the spectral shape increases exponentially towards the EUV

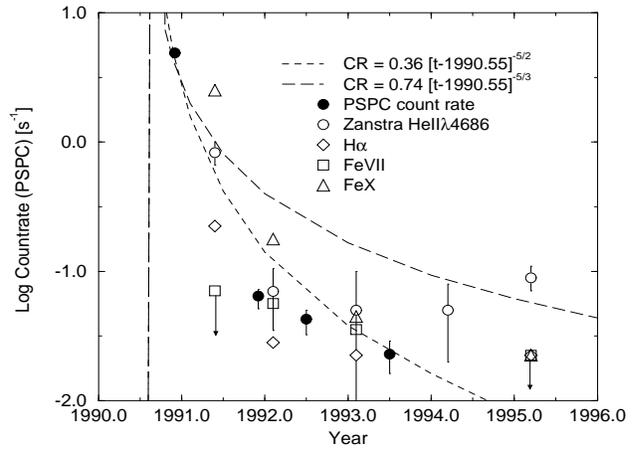

**Fig. 4.** X-ray lightcurve of IC3599, compared with X-ray fluxes inferred by the Zanstra method from HeIIλ4686. Dashed lines show lightcurves $\propto [t - t_0]^{-s}$. Lightcurves with $t^{-s}$ asymptotes where $s = 5/2$ (Rees 1988) or, more probably, $s = 5/3$ (Rees 1990) have been predicted as an indication for the tidal disruption of a star by a $10^6 M_\odot$ black hole

*Acknowledgements.* We thank T. Boller, H.L. Marshall, W.N. Brandt for valuable discussions.